\documentclass[prd,superscriptaddress,showpacs,nofootinbib,amsmath,amssymb,aps,10pt]{revtex4}

\usepackage{bm}
\usepackage{amsfonts}
\usepackage{latexsym}
\usepackage[latin1]{inputenc}
\usepackage{graphicx}
\usepackage{amsmath}
\usepackage{palatino}
\usepackage{mathpazo}
\usepackage{epstopdf}
\usepackage{booktabs}
\usepackage{dcolumn}
\usepackage{array}

\usepackage{wrapfig}
\usepackage{wasysym}

\def\jnl@style{\it}
\def\aaref@jnl#1{{\jnl@style#1}}

\def\aaref@jnl#1{{\jnl@style#1}}

\def\aj{\aaref@jnl{AJ}}                   
\def\apj{\aaref@jnl{ApJ}}                 
\def\apjl{\aaref@jnl{ApJ}}                
\def\apjs{\aaref@jnl{ApJS}}               
\def\apss{\aaref@jnl{Ap\&SS}}             
\def\aap{\aaref@jnl{A\&A}}                
\def\aapr{\aaref@jnl{A\&A~Rev.}}          
\def\aaps{\aaref@jnl{A\&AS}}              
\def\mnras{\aaref@jnl{Mon.~Not.~Roy.~Astron.~Soc.}}             
\def\prd{\aaref@jnl{Phys.~Rev.~D}}        
\def\prc{\aaref@jnl{Phys.~Rev.~C}}  
\def\prl{\aaref@jnl{Phys.~Rev.~Lett.}}    
\def\qjras{\aaref@jnl{QJRAS}}             
\def\skytel{\aaref@jnl{S\&T}}             
\def\ssr{\aaref@jnl{Space~Sci.~Rev.}}     
\def\zap{\aaref@jnl{ZAp}}                 
\def\nat{\aaref@jnl{Nature}}              
\def\aplett{\aaref@jnl{Astrophys.~Lett.}} 
\def\apspr{\aaref@jnl{Astrophys.~Space~Phys.~Res.}} 
\def\physrep{\aaref@jnl{Phys.~Rep.}}      
\def\physscr{\aaref@jnl{Phys.~Scr}}       
\def\commat{\aaref@jnl{Comm.~Math.~Phys.}}              
\def\science{\aaref@jnl{Science}}               
\def\cqg{\aaref@jnl{Classical Quant.~Grav.}}            
\def\jpcs{\aaref@jnl{JPCS}}                                     
\def\ijmpd{\aaref@jnl{Int.~J.~Mod.~Phys.~D}}                    
\def\grg{\aaref@jnl{Gen.~Relat.~Gravit.}}               
\def\rpp{\aaref@jnl{Rep.~Prog.~Phys.}}          
\def\npa{\aaref@jnl{Nucl.~Phys.~A}}        
\def\lrr{\aaref@jnl{Living Rev.~Rel.}}                   
\def\jcap{\aaref@jnl{J.~Cosmology Astropart.~Phys.}}    
\def\rmp{\aaref@jnl{Rev.~Mod.~Phys.}}   

\allowdisplaybreaks[1]

\begin{document}

\title{Orbital and epicyclic frequencies in massive scalar-tensor theory with self-interaction }

\author{Kalin V. Staykov}
\email{kstaykov@phys.uni-sofia.bg}
\affiliation{Department of Theoretical Physics, Faculty of Physics, Sofia University, Sofia 1164, Bulgaria}

\author{Daniela D. Doneva}
\email{daniela.doneva@uni-tuebingen.de}
\affiliation{Theoretical Astrophysics, Eberhard Karls University of T\"ubingen, T\"ubingen 72076, Germany}
\affiliation{INRNE - Bulgarian Academy of Sciences, 1784  Sofia, Bulgaria}

\author{Stoytcho S. Yazadjiev}
\email{yazad@phys.uni-sofia.bg}
\affiliation{Department of Theoretical Physics, Faculty of Physics, Sofia University, Sofia 1164, Bulgaria}
\affiliation{Institute of Mathematics and Informatics, Bulgarian Academy of Sciences, Acad. G. Bonchev Street 8, Sofia 1113, Bulgaria}


\begin{abstract}
 Testing modified theories of gravity with direct observations of the parameters of a neutron star is not the optimal way of testing gravitational theories. However, observing electromagnetic signals originating from the close vicinity of the compact object my turn out an excellent way of probing spacetime in strong field regime. A promising candidate for doing so are the so-called quasi-periodic oscillations, observed in the X-ray light curves of some pulsars. Although the origin of those oscillations is unknown, one thing most of the models describing them have in common is that in one way or another they incorporate the radius of the innermost stable circular obit, and the orbital and the epicyclic frequencies of particles moving around the compact object. In this paper we study the aforementioned quantities in the context of massive scalar-tensor theory and massive scalar-tensor theory with self-interaction, both of which in strong regime allow for significant deviations from General relativity for values for the  free parameters of the theory in correlation with the observations.   
\end{abstract}

\maketitle
\date{}

\section{Introduction}

Ones of the simplest and well-motivated modified theories of gravity are scalar-tensor
theories (STTs) of gravity, in which gravity is mediated by the spacetime metric tensor and a dynamical scalar field. 
In the past decades the massless STT with coupling function of the form $\alpha = \beta\varphi$ (where $\varphi$ is the scalar field and $\beta$ is free parameter) was among the most popular ones. This theory gained its popularity due to the fact that in weak field regime its predictions coincide with GR but in strong field regime significant deviations could be observed. However, in the recent years astrophysical observations of binary pulsars set tight constraints on the allowed values for the free parameter $\beta$ in the theory, which made it more ot less indistinguishable from GR. Recently it was shown in \cite{Ramazanoglu2016,Popchev2015,Yazadjiev2016} that this can be overcome if a potential with a massive term is added to the Lagrangian of the theory. The massive term significantly increases the interval of the allowed values for the  parameter $\beta$ and in the same time observational constraints on the mass of the field can be set as well. In those papers the authors showed that the presence of the mass term suppresses the spontaneous scalarization, but still significant deviations from GR can be observed for values  of the parameters in agreement  with the observations. Recently those studies were extended in \cite{Staykov2018, Popchev2018} by adding a self-interaction quartic therm in the potential. However, there are no observational constraints on the self-interaction constant. The presence of the self-interaction term in the potential additionally  suppresses the scalarization of the neutron star models, but in this case as well significant deviations from GR can be observed for values for the parameters in correlation with observations. 

The recent detection of gravitational waves \cite{Collaboration2018}, even more the neutron star -- neutrons star merger \cite{Abbott2017} with its multi-messenger  detection made even more important the proper study of modified theories in both gravitational wave and in the electromagnetic spectrum. One important, but still unexplained phenomena in the electromagnetic spectrum are the so-called quasi-periodic oscillations (QPO) observed in the X-ray light curves of some pulsars. The QPOs are Hz to kHz oscillations in the X-ray flux of compact object (neutrons stars and black hole candidates). The kHz QPOs are supposed to originate from the inner edge of the accretion disk, which means that they may turn out to be excellent probes for the strong field regime in the vicinity of the compact objects. 

A good estimate how close to the star the source of the signal can be is the so-called innermost stable circular orbit (ISCO). The ISCO orbit in many models is supposed to be the inner edge of the accretion disc. On the other hand, ISCO may turn out to be important for the compact object mergers due to the fact that after ISCO the two bodies should start falling rapidly to each other. The QPO origin is unknown and different models based on different mechanism explaining them with variable success -- one can see \cite{Klis2006} for comprehensive review.  In one way or another most QPO models incorporate the orbital frequency, and the orbital and the epicyclic frequency  of a particle on a circular orbit. In general there are two main classes of models based on the different  mechanisms behind the QPOs. The first one is based on orbital and epicyclic motion of matter around the central object (for example \cite{Miller1998,Stella1999,Stella2001, Abramowicz2004,Pappas2012b, Motta2014, Pappas2015c, Maselli2015, Staykov2015a}), and the second one is based on oscillations and instabilities in an accretion disc around the compact object (for example \cite{Rezzolla2003, Rezzolla2003a, Montero2004, Fragile2016, Avellar2018}). 

This paper is structured as follow. In section II we present the mathematical basics. In the first part of that section we preset the field equations for constructing the neutron star model and the background solution in general.  In the second part of section II we present the general scheme for deriving the orbital on the epicyclic frequencies of particle on circular orbit in stationary and axisymmetric spacetime, as well as the conditions for finding the ISCO orbit. In section III we present and discuss the numerical results for massive scalar-tenros theory and massive scalar-tensor theory with self-interaction. The paper ends with a Conclusion. 

\section{Basic equations} 

\subsection{The background solution}

For simplicity, the mathematical part of this paper is in the more convenient  Einstein frame, but all presented results in the following section are in the physical Jordan frame. The Einstein frame STT action is 

\begin{eqnarray}\label{EFA}
S=\frac{1}{16\pi G} \int d^4x \sqrt{-g_{*}}\left[ R_{*} - 2
g_{*}^{\mu\nu}\partial_{\mu}\varphi \partial_{\nu}\varphi - V(\varphi)
\right] + S_{\rm
	matter}(A^2(\varphi)g^{*}_{\mu\nu},\chi),
\end{eqnarray}
where $R_{*}$ is the Ricci scalar curvature with respect to the Einstein frame metric $g^{*}_{\mu\nu}$. In Einstein frame, the scalar-tensor theories are specified by the function $A(\varphi)$, which gives the conformal transformation of the metric between both frames, and the scalar-field potential   $V(\varphi)$. In the present paper we will extend the study in \cite{Yazadjiev2016, Staykov2018}, and therefore adopt the conformal factor and the potential used in those papers, namely 
\begin{equation}
A(\varphi)=e^{\frac{1}{2}\beta \varphi^2}
\end{equation} 
 and    
\begin{equation}
V(\varphi)=2m^2_{\varphi}\varphi^2 + \lambda \varphi^4.
\end{equation}
This potential is the simplest one with forth order self-interaction term. The first term in the potential $V(\varphi)$ is the standard massive term, considered in previous studies of massive STT \cite{Popchev2015,Ramazanoglu2016,Yazadjiev2016,Doneva:2016xmf} while the second term describes self-interaction of the scalar field \cite{Staykov2018, Popchev2018}. 

The Jordan and the Einstein frame metrics, $ g_{\mu\nu}$ and $g^{*}_{\mu\nu}$, are connected via conformal transformation  $g_{\mu\nu}= A^2(\varphi)g^{*}_{\mu\nu}$  and the gravitational scalar respectively by $\Phi=A^{-2}(\varphi)$. The energy-momentum tensor transformation between both frames is given by the relation $T^{*}_{\mu\nu}=A^2(\varphi)T_{\mu\nu}$, where  $T^{*}_{\mu\nu}$  and  $T_{\mu\nu}$ are the Einstein and the Jordan frame ones, respectively.
For a perfect fluid, as it is modeled to constitute the interior of the star, the relations between the energy density and pressure in both frames are given by $\rho_{*}=A^4(\varphi) \rho$ and $p_{*}=A^4(\varphi)p$.

In this paper we are using slow rotation approximation in first order in the angular velocity $\Omega$, i.e. keeping only first order terms. This approximation is suitable for the purpose of this paper, because it allow us to study with good accuracy models rotting with frequency up to about $f = 160$ Hz, which covers the majority of the observed pulsars. In addition we consider stationary and axisymmetric spacetime as well as stationary and axisymmetric scalar field and  fluid configurations. The Einstein frame spacetime metric in this case has the form \cite{Hartle1967}

\begin{eqnarray}
ds_{*}^2= - e^{2\phi(r)}dt^2 + e^{2\Lambda(r)}dr^2 + r^2(d\theta^2 +
\sin^2\theta d\vartheta^2 ) - 2\omega(r,\theta)r^2 \sin^2\theta  d\vartheta dt.
\end{eqnarray}

This is possible, because the metric function $\omega$ is in linear order of $\Omega$,  but the rotational corrections to other metric functions, the scalar field, the fluid energy density and pressure are of order ${\cal O}(\Omega^2)$. 

The slow rotation approximation dimensionally reduced Einstein frame field equations, derived from the action (\ref{EFA}) are the following

\begin{eqnarray} \label{eq:FieldEq}
&&\frac{1}{r^2}\frac{d}{dr}\left[r(1- e^{-2\Lambda})\right]= 8\pi G
A^4(\varphi) \rho + e^{-2\Lambda}\left(\frac{d\varphi}{dr}\right)^2
+ \frac{1}{2} V(\varphi), \nonumber\\
&&\frac{2}{r}e^{-2\Lambda} \frac{d\phi}{dr} - \frac{1}{r^2}(1-
e^{-2\Lambda})= 8\pi G A^4(\varphi)  p +
e^{-2\Lambda}\left(\frac{d\varphi}{dr}\right)^2 - \frac{1}{2}
V(\varphi),\nonumber\\
&&\frac{d^2\varphi}{dr^2} + \left(\frac{d\phi}{dr} -
\frac{d\Lambda}{dr} + \frac{2}{r} \right)\frac{d\varphi}{dr}= 4\pi G
\alpha(\varphi)A^4(\varphi)( \rho-3 p)e^{2\Lambda} + \frac{1}{4}
\frac{dV(\varphi)}{d\varphi} e^{2\Lambda}, \\
&&\frac{d p}{dr}= - ( \rho +  p) \left(\frac{d\phi}{dr} +
\alpha(\varphi)\frac{d\varphi}{dr} \right), \nonumber\\
&&\frac{e^{\Phi-\Lambda}}{r^4} \partial_{r}\left[e^{-(\Phi + \Lambda)} r^4 \partial_{r}{\bar\omega} \right]  + \frac{1}{r^2\sin^3\theta} \partial_{\theta}\left[\sin^3\theta\partial_{\theta}\bar\omega \right]= 16\pi GA^4(\varphi)( \rho + p)\bar\omega , \nonumber
\end{eqnarray}
where the function $\bar\omega$ is defined as $\bar\omega = \Omega - \omega$, and the coupling function $\alpha(\varphi)$ is defined by $\alpha(\varphi)=\frac{d\ln A(\varphi)}{d\varphi}$.

The above system of equations (\ref{eq:FieldEq}),  supplemented with the equation of state for the the stellar matter and the appropriate boundary conditions, describes the interior  and the exterior of the neutron star, and it is used for deriving the background solutions used in this study. The exterior space-time of the neutron star is described by the system (\ref{eq:FieldEq}), by setting $\rho=p=0$.

At the center of the star we have the natural boundary conditions -  $\rho(0)=\rho_{c},  \Lambda(0)=0, {\rm and} \frac{d{\varphi}}{dr}(0)= 0$, where $\rho_{c}$ is the constant central density. From the requirement for asymptotic flatness, at infinity we have $\lim_{r\to \infty}\phi(r)=0, \lim_{r\to \infty}\varphi (r)=0$ (see e.g. \cite{Yazadjiev2014}). The coordinate radius $r_S$ of the star in the Einstein frame is determined by the standard condition $p(r_S)=0$, and the physical radius is obtained by conformal transformation in the Jordan frame - $R_{S}= A[\varphi(r_S)] r_S$.

The slow rotation approximation allow us to separate the equation for $\bar \omega$  from the other equations in the system (\ref{eq:FieldEq}) and it can be considerably simplified. The simplification procedure one can find explained in \cite{Yazadjiev2016}, and it leads to $\bar{\omega}$ as function of $r$ only, and the equation transforms into

\begin{eqnarray}\label{OR}
\frac{e^{\Phi-\Lambda}}{r^4} \frac{d}{dr}\left[e^{-(\Phi+ \Lambda)}r^4 \frac{d{\bar\omega}(r)}{dr} \right] =
16\pi G A^4(\varphi)(\rho + p){\bar\omega}(r).
\end{eqnarray}
The natural boundary condition for ${\bar\omega}$ to ensure its regularity at the center of the star is $\frac{d{\bar\omega}}{dr}(0)= 0$, and at infinity $\lim_{r\to \infty}{\bar\omega}=\Omega$.

For the numerical computations and in the presentation of the results in the next section we are using dimensionally reduced parameters 
 $m_{\varphi}\to m_{\varphi} R_{0}$ and $\lambda \to \lambda R_{0}^2$, where  $R_{0}=1.47664 \,{\rm km}$ is one half of the solar gravitational radius.

\subsection{ISCO, orbital and epicyclic frequencies}

We continue our discussion with brief presentation of the basic steps in the derivation of  the equations for the radius of the innermost stable circular orbit (ISCO), the equations for the radial and for the vertical epicyclic frequencies and for the orbital frequency  \cite{Ryan1995,Maselli2015,Shibata1998,Pappas2012}. The equations describing stable neutron star models used as a background solution were discussed in the first part of this section, and they are thoroughly studied in \cite{Staykov2018}.

We are considering the most general form for a stationary and axisymmetric spacetime metric in the Jordan frame

\begin{eqnarray} \label{Metric}
ds^2 = g _{tt}dt^2 + g_{rr}dr^2 + g_{\theta \theta}d\theta^2 + 2g_{t\vartheta}dtd\vartheta + g_{\vartheta\vartheta}d\vartheta^2,
\end{eqnarray}
where all the metric functions depend only on the coordinates $r$ and $\theta$. 

Massive particles in gravitational field only move on
timelike geodesics of the Jordan frame metric (\ref{Metric}). The stationary and axial Killing symmetries of metric, generated by the Killing vectors $\frac{\partial}{\partial t}$ and
$\frac{\partial}{\partial \vartheta}$, give rise to two constants of motion, namely $E=-u_t$ which corresponds to the energy per unit mass and $L=u_{\vartheta}$ which corresponds to the angular momentum per unit mass. $u^{\mu} = \dot{x}^{\mu} = dx^{\mu}/d\tau$ is the four-velocity of the particle. One can show without any difficulty that the conservation laws can be rewritten in the form
\begin{eqnarray}
&\frac{dt}{d\tau} = \frac{Eg_{\vartheta\vartheta} + Lg_{t\vartheta}}{g2}, \\
&\frac{d\vartheta}{d\tau} = -\frac{Eg_{t\vartheta} + Lg_{tt}}{g2},
\end{eqnarray}
where  $g2 = g_{t\vartheta}^2 - g_{tt}g_{\vartheta\vartheta}$ is defined for simplicity. The normalization condition  $g^{\mu\nu} u_{\mu}u_{\nu} = -1$ for the four-velocity give us

\begin{eqnarray} \label{eq:4v_norm}
g_{rr}\dot{r}^2 + g_{\theta\theta}\dot{\theta}^2 + E^2U(r,\theta) = -1,
\end{eqnarray}
where we have defined

\begin{eqnarray}
U(r,\theta)= \frac{g_{\vartheta\vartheta} + 2l g_{t\vartheta} + l^2 g_{tt}}{g2},
\end{eqnarray}
and $l=L/E$ is the proper angular momentum.

In the equatorial plane, $\theta=\frac{\pi}{2}$, the problem reduces to an one dimensional problem with an effective equation of motion 

\begin{eqnarray}
\dot{r}^2 = V(r),
\end{eqnarray}
and effective potential

\begin{eqnarray}
V(r)=g_{rr}^{-1}\left[-1 - E^2 U(r,\theta=\frac{\pi}{2})\right]. 
\end{eqnarray}

For given $E$ and $L$ of the particle, the stable circular orbit with a radius $r_c$ is determined by the conditions $V(r_c)= 0 = V^{'}(r_c)$ and $V^{\prime\prime}(r_c)>0$, where with prime we denote the derivative with respect to $r$, and the radius of ISCO is given by the vanishing second derivative of the potential --  $V^{\prime\prime}(r_c)=0$. The angular velocity $\Omega_p$ of a particle moving on a circular orbit in the equatorial plane 
can be found from  the geodesic equation written in the form 

\begin{eqnarray}
\frac{d}{d\tau}\left(g_{\mu\nu}\frac{dx^\nu}{d\tau}\right)= \frac{1}{2}\partial_{\mu}g_{\nu\sigma} \frac{dx^\nu}{d\tau} \frac{dx^\sigma}{d\tau}.
\end{eqnarray}   

For the radial coordinate this equation translates into  

\begin{eqnarray}
\partial_{r}g_{tt} (\frac{dt}{d\tau})^2 + 2 \partial_{r}g_{t\vartheta} \frac{dt}{d\tau} \frac{d\vartheta}{d\tau} + \partial_{r}g_{\vartheta\vartheta}(\frac{d\vartheta}{d\tau})^2=0 ,
\end{eqnarray} 
from which by taking into account the definition for  the angular velocity  $\Omega_{p}=\frac{u^\vartheta}{u^{t}}=\frac{d\vartheta}{dt}$ we obtain 

\begin{eqnarray}
\Omega_p = \frac{d\vartheta}{dt} = \frac{-\partial_{r} g_{t\vartheta} \pm \sqrt{(\partial_{r}g_{t\vartheta})^2 - \partial_{r}g_{tt}\partial_{r}g_{\vartheta\vartheta}}}{\partial_{r}g_{\vartheta\vartheta}}.
\end{eqnarray} 

The positive sign in the above equation corresponds to prograde orbits and the negative sign to retrograde ones. In the present work, we are considering only prograde orbits. 

To derive the epicyclic frequencies one should investigate small perturbations in radial and in vertical direction of a stable orbit. 
The radial and the vertical perturbations we write in the form

\begin{eqnarray} \label{eq:pert_orb}
r(t) = r_c + \delta r(t), \quad \theta (t) = \frac{\pi}{2} + \delta \theta(t),
\end{eqnarray}
where $\delta r(t)$ and $\delta \theta(t)$ are the perturbations to the stable circular orbit with coordinate radius $r_c$ in the equatorial plane ($\theta = \pi/2$). The perturbations could be written explicitly in the form $\delta r(t) \sim e^{2\pi i \nu_{r}t} $ and $\delta \theta(t) \sim   e^{2\pi i \nu_{\theta}t} $. By substituting (\ref{eq:pert_orb}) into eq. (\ref{eq:4v_norm}) one can obtain the expressions for the radial and for the vertical epicyclic frequencies:

\begin{eqnarray}
\nu_{r}^{2} = \frac{\left(g_{tt} + \Omega_p g_{t\vartheta}\right)^2}{2\left(2\pi\right)^2 g_{rr}} \partial_{r}^{2} U\left(r_c,\frac{\pi}{2}\right) \\
\nu_{\theta}^{2} = \frac{\left(g_{tt} + \Omega_p g_{t\vartheta}\right)^2}{2\left(2\pi\right)^2 g_{\theta\theta}} \partial_{\theta}^{2} U\left(r_c,\frac{\pi}{2}\right).
\end{eqnarray}

For static neutron stars, $f = 0$, the orbital frequency and the vertical epicyclic frequency coincide, i.e. $\nu_{\theta} = \nu_p$.
At ISCO the radial epicyclic frequency is equal to zero, and for smaller radius it is negative, which shows a radial instabilities for orbits with radius smaller than ISCO.

\section{Numerical results}

The constraints   for the parameter $\beta$ and for the mass of the scalar field $m_{\varphi}$ and the reasoning for them have already been thoroughly discussed in \cite{Ramazanoglu2016,Yazadjiev2016,Staykov2018} and we will not repeat this discussion here, but will briefly present only the constrained intervals of allowed values. For the mass of the scalar field $m_{\varphi}$

\begin{equation}
10^{-16} {\rm eV} \lesssim m_{\varphi} \lesssim 10^{-9} {\rm eV},
\end{equation}  
which roughly corresponds to $10^{-6} \lesssim m_{\varphi} \lesssim 10$ in our dimensionless units.

For sufficiently large mass of the scalar field, the interval of allowed values for the parameter $\beta$ is $3 \lesssim -\beta \lesssim 10^{3}$, which is significantly wider compared to the massless case. As long as the self-interaction constant $\lambda$ is concerned, the only constraint that we have is that it should be positive in order for the potential to be positive, and we restrict ourself to the values used in  \cite{Staykov2018}.

In this study we used two popular EOS with maximal mass in GR in correlation with the observations, the so-called APR4 EOS \cite{AkmalPR}, and the so-called SLy EOS \cite{Douchin2001}, and for both the piecewise polytropic approximation is used \cite{Read2009}. However, due to the qualitatively identical behavior we chose to present only the results for EOS SLy. The reasoning for using only one EOS is the same as in \cite{Staykov2018} -- the system (\ref{eq:FieldEq}) has a three parameter ($\beta$, $m_\varphi$ and $\lambda$) family of solutions, and including multiple EOS will only make the results overwhelming. 
As we have mentioned we are using the slow rotation approximation. And we study three different rates of rotation -- static models with $f = 0$ Hz, models with $f = 80$ Hz, and models with $f=160$ Hz, where $f$ is the rotational frequency of the star ($f = \frac{\Omega}{2 \pi}$). However, due to the small effect of the rotation on the studied parameters, in most of the cases we will present only the static case results, and the effect of the rotation will only be discussed briefly.

\subsection{Massive scalar-tensor theory}

\begin{figure}[]
	\centering
	\includegraphics[width=0.45\textwidth]{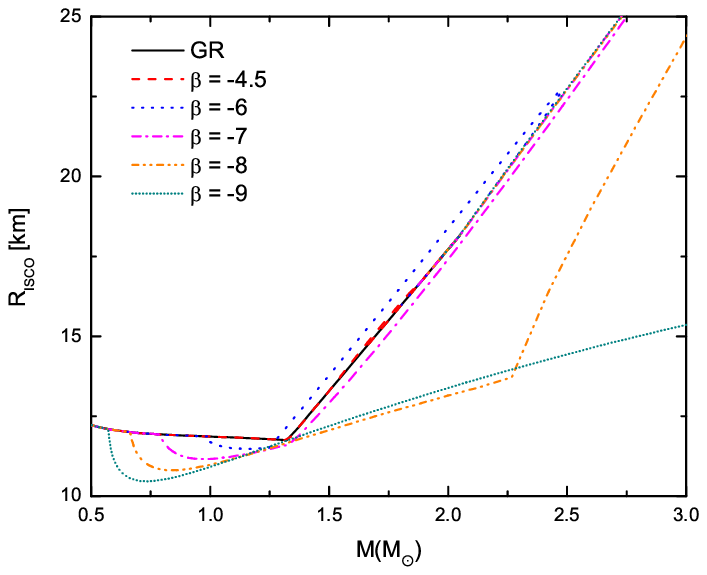}
	\includegraphics[width=0.45\textwidth]{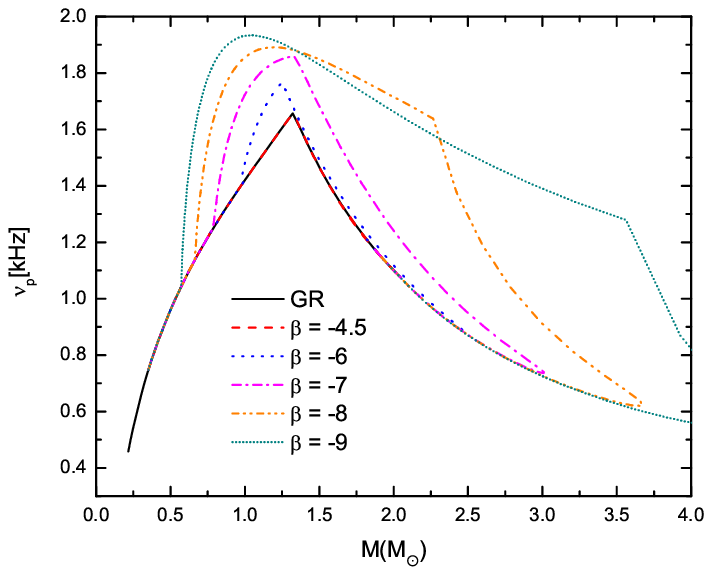}
	\caption{Left: The radius of ISCO as function of the mass of the models in massless STT with different values for the parameter $\beta$ (in different patterns and colours). Right: The orbital frequency as function of the mass of the models in massless STT. The same notations as in the left panel are used. }
	\label{Fig:ISCO_nup_b}
\end{figure}

\begin{figure}[]
	\centering
	\includegraphics[width=0.45\textwidth]{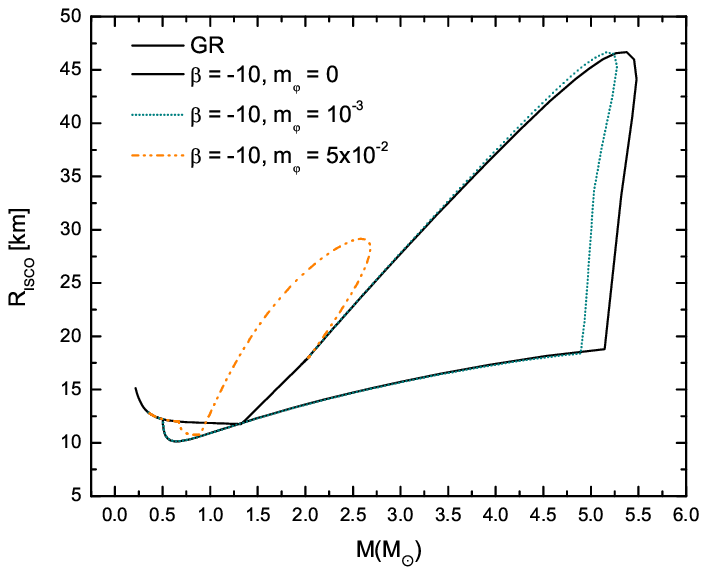}
	\includegraphics[width=0.45\textwidth]{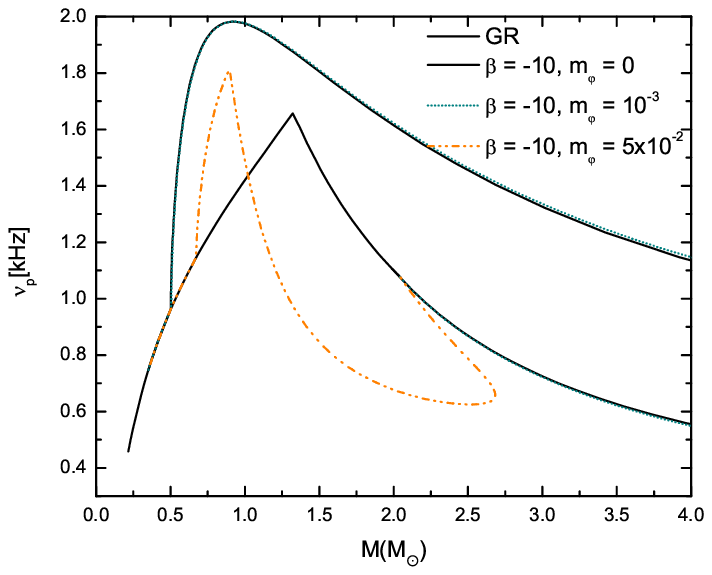}
	\caption{ Left: The radius of ISCO as function of the mass of the models in massive STT with different values for $\beta = -10$ and different values for the mass of the field $m_{\varphi}$ (in different patterns and colours). Right: The orbital frequency as function of the mass of the models in massive STT. The values for the parameters and the notations are the same as in the left panel.  }
	\label{Fig:ISCO_nup_m}
\end{figure}

We start our study with the simplest case of STT, and gradually extend it to the the theory which is of main interest for us, namely the massive STT with self interaction. For better understanding of the effect of the parameters in the latter theory in Fig. \ref{Fig:ISCO_nup_b} and Fig. \ref{Fig:ISCO_nup_m} we present the radius of ISCO and the orbital frequency in massless STT and in massive STT correspondingly. Although the massive STT is a viable alternative to GR \cite{Ramazanoglu2016, Yazadjiev2016, Staykov2018} we will discus it shortly in order to concentrate more on the self-interacting case. 

In Fig. \ref{Fig:ISCO_nup_b} we study the massless case for different values for the parameter $\beta$. The used values for the parameter are quite small compared to the observational constraints. We chose them in such way  regardless the fact they are ruled-out by the observation, because they give us the limiting case obtained from the massive theory for vanishing mass of the scalar field. In the left panel we plot the radius of ISCO as function of the mass of the star (if all circular orbits in the exterior space are stable, we plot the radius of the star). It is interesting to point out, that with the decrease of the parameter $\beta$ the mass at which ISCO appears increases significantly compared to the GR one. When ISCO appears, for low values for $\beta$ it's radius rapidly increases for quite short interval of masses, and for big values for $\beta$ gradually increases for wider interval of masses, and then it starts to converge to the GR case.   For high values for $\beta$ ISCO appears for masses lower compared to the GR one, and that mass increases significantly above the GR one when the values for $\beta$ decrease.  In the right panel, we plot the orbital frequency as function of the mass (if all circular orbits are stable, we calculate the orbital frequency at the surface of the star). It is clearly visible at which point ISCO appears - there is a sharp edge in the graph, and after it, the frequency rapidly decreases.  

In Fig. \ref{Fig:ISCO_nup_m} we study ISCO and the orbital frequency for the smallest value for the parameter $\beta$ we are using in this study, and some values for the mass of the scalar field. One can see that for high mass of the field, ISCO appears for low mass stars, and it allows for quite significant deviations from GR (both in ISCO and in the orbital frequency) for the same interval of masses. For the presented in the figure values for the parameters, the radius of ISCO increases with more than 60\%, compared to GR, and the orbital frequency decreases with more than 40\% compared to GR. 
We find these models very interesting because they give us significant deviations between modes with ISCO in both theories for the same mass interval (as in GR). This is not the case for the massless case or the massive case with low values for $\beta$ and low mass of the field.

\subsection{Massive scalar-tensor theory with self-interaction}

In this section we extend our study to the case we are most interested in -- the massive STT with self-interaction. We will concentrate on two values for the parameter $\beta$ and different values for the mass of the field and for the self-interaction constant. 

\begin{figure}[]
	\centering
	\includegraphics[width=0.45\textwidth]{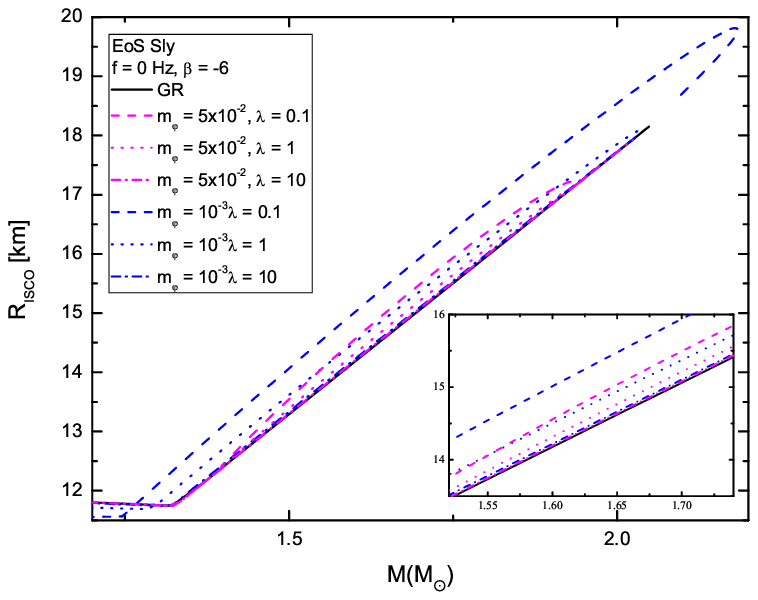}
	\includegraphics[width=0.45\textwidth]{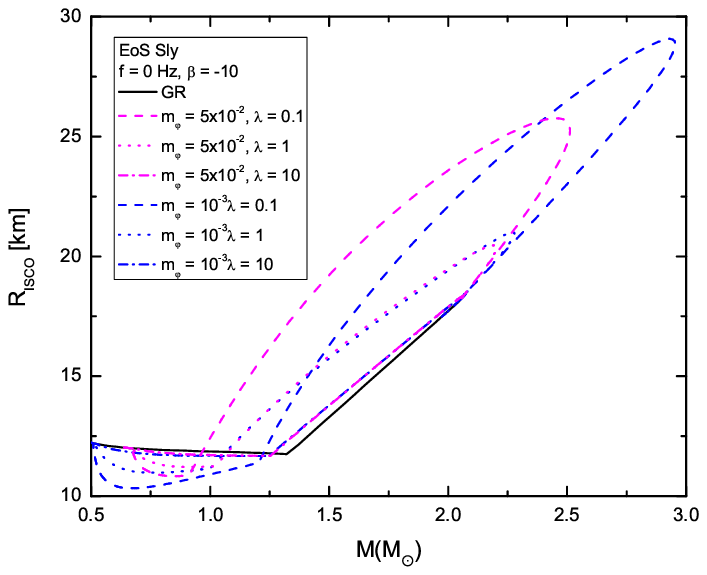}
	\caption{ The radius of ISCO as function of the mass of the models in massive STT with self-interaction for different combinations of the mass of the field $m_{\varphi}$ and the self-interaction parameter $\lambda$ (in different patterns and colours). Left: $\beta = -6$ and right: $\beta = -10$.  }
	\label{Fig:ISCO}
\end{figure}

In Fig. \ref{Fig:ISCO} we plot the radius of ISCO as function of the mass of the star. In the left panel we plot models with $\beta = -6$, and in the right -- $\beta = -10$. As expected, the deviations in the left panel are quite smaller (less than 6\%), and in the right panel, due to the higher value for $\beta$, the deviations are more significant (up to 40\%). As one can expect knowing the results in \cite{Staykov2018}, with the increase of the mass of the field, the self-interaction constant, or both, the results converge to the GR ones. However, it is very interesting to point out that in the right panel ($\beta = -10$) the results does not look like the deviation behave monotonically with the parameters -- for higher values of the mass of the field one observes higher deviations from GR compared to models with lower values for the mass of the field.  As it appears, this can be explained by the fact that for the higher mass of the field, ISCO appears for models with lower mass compared to the low scalar field mass case.

\begin{figure}[]
	\centering
	\includegraphics[width=0.45\textwidth]{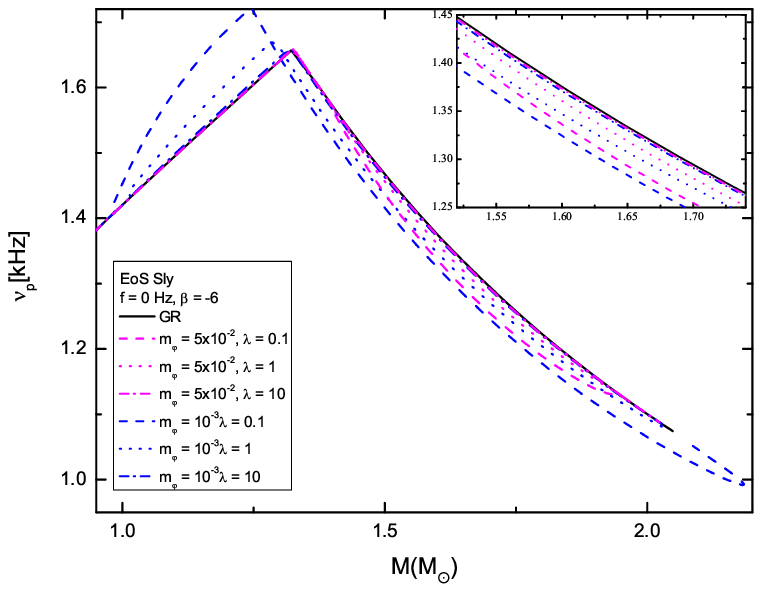}
	\includegraphics[width=0.45\textwidth]{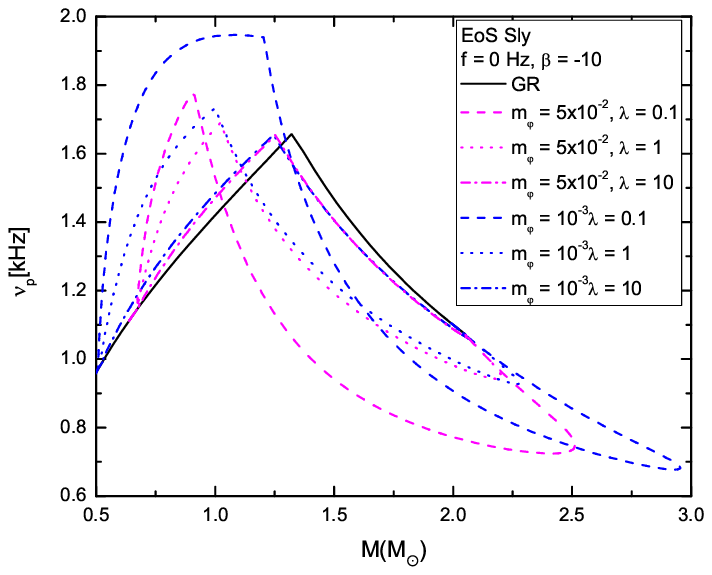}
	\caption{The orbital frequency as function of the mass of the models in massive STT with self-interaction for different combinations of the mass of the field $m_{\varphi}$ and the self-interaction parameter $\lambda$ (in different patterns and colours). Left: $\beta = -6$ and right: $\beta = -10$.}
	\label{Fig:nu_p}
\end{figure}

In Fig. \ref{Fig:nu_p} we plot the orbital frequency as function of the mass of the models. Models at which ISCO appears are distinguishable from the rest by the sharp edge on the graphs.  In both panels one can see  that for all combinations of parameters (no matter if they give small or high deviation from GR) if all orbits in the exterior are stable, hence the orbital frequency is calculated at the surface of the star, the frequency is higher, compared to the GR one. For models, for which ISCO exists, however, the frequency is always lower, compared to the GR one. As one can expect by the results for ISCO, for beta $\beta = -6$ the frequency increases with the increase of the mass of the filed, as well as with the increase of the coupling constant $\lambda$. The maximal deviation is about 4\% in this case. For $\beta = -10$ the models with higher mass of the field show higher deviation (up to about 35 \%) than the models with low mass of the field. In this case as well the results converge to GR with the increase of $\lambda$.

\begin{figure}[]
	\centering
	\includegraphics[width=0.45\textwidth]{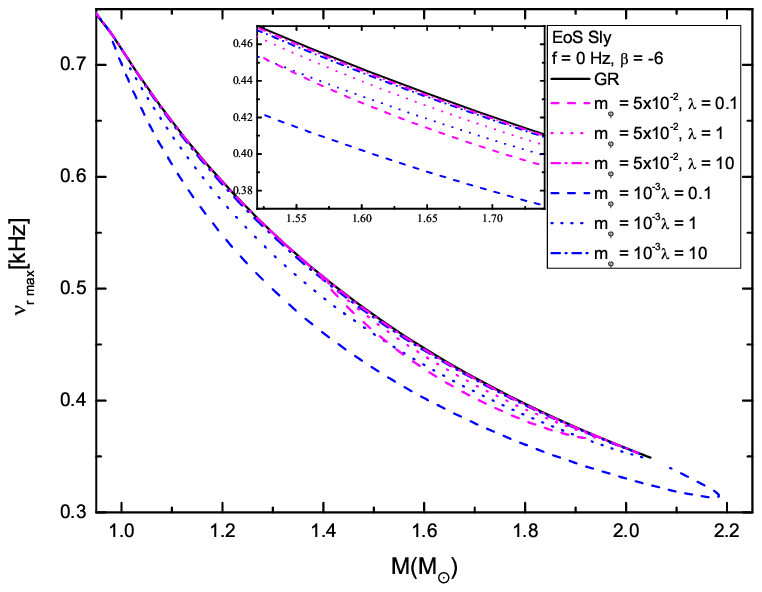}
	\includegraphics[width=0.45\textwidth]{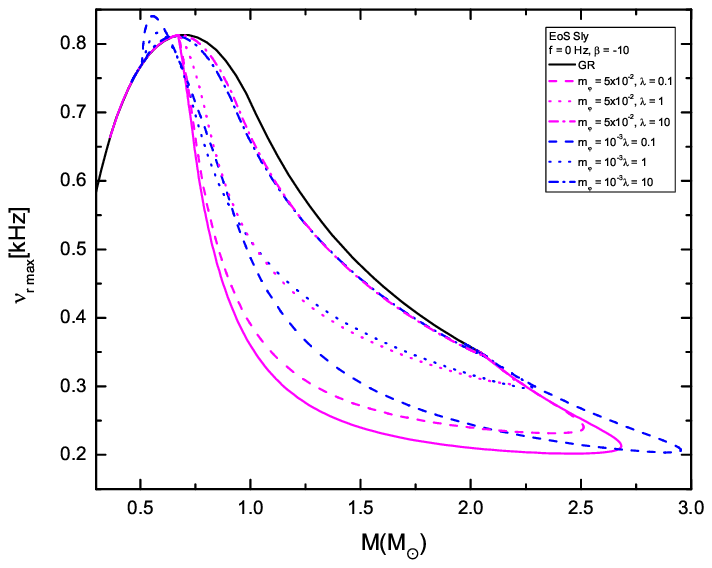}
	\caption{The maximal radial epicyclic frequency as function of the mass of the models in massive STT with self-interaction for different combinations of the mass of the field $m_{\varphi}$ and the self-interaction parameter $\lambda$ (in different patterns and colours). Left: $\beta = -6$ and right: $\beta = -10$.}
	\label{Fig:nu_r_max}
\end{figure}

In Fig. \ref{Fig:nu_r_max} we continue our study with the maximal radial frequency. If the radial frequency does not have a maximum outside of the star, we plot the frequency at it's surface. In this case as well the frequency is always lower compared to the GR one (with maximal deviations about 10\% for $\beta = -6$ and 50\% for$\beta = -10$). One can see that this quantity shows the already discussed behavior with the mass of the field.   

\begin{figure}[]
	\centering
	\includegraphics[width=0.45\textwidth]{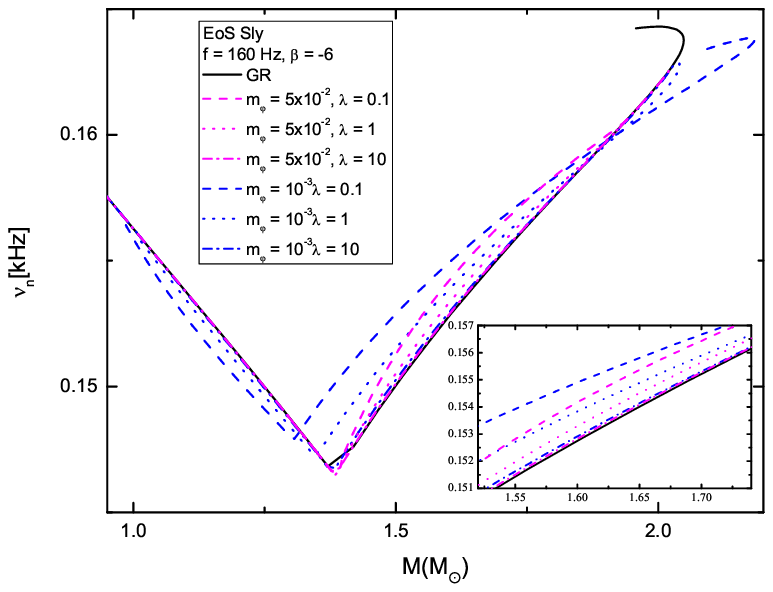}
	\includegraphics[width=0.45\textwidth]{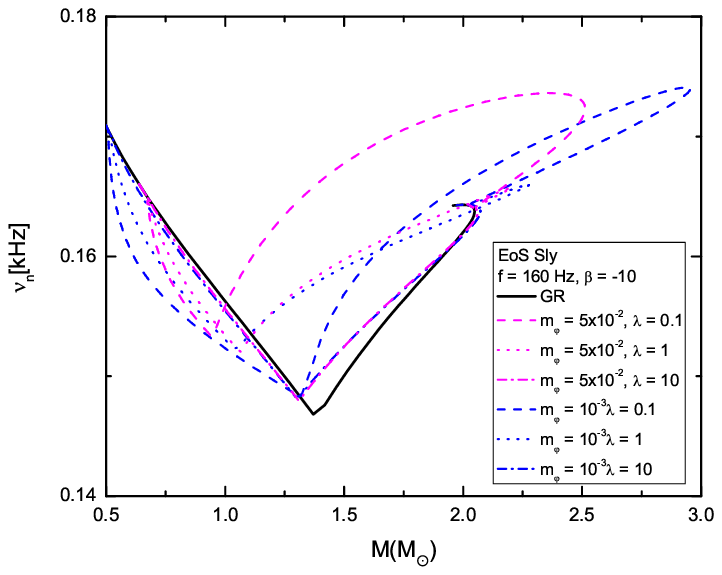}
	\caption{The nodal precession frequency as function of the mass of the models in massive STT with self-interaction for different combinations of the mass of the field $m_{\varphi}$ and the self-interaction parameter $\lambda$ (in different patterns and colours). Left: $\beta = -6$ and right: $\beta = -10$.}
	\label{Fig:nu_n}
\end{figure}

In Fig. \ref{Fig:nu_n} we plot the nodal precession frequency as a function of the mass of the models for two different values of parameters $\beta$ and different combinations between the mass of the scalar field $m_{\varphi}$ and the self-interaction constant $\lambda$. If all circular orbits are stable, we calculate the frequency at the surface of the star, and in the other case - on ISCO. Both regimes are clearly distinguishable by the sharp edge in the graphs. Due to the fact that the nodal precession frequency is given by the difference between the orbital frequency and the vertical epicyclic one ($\nu_n = \nu_p - \nu_{\theta}$), which coincide in the static case, in this figure we plot the highest rate of rotation we studied (which is the in the upper boundary of frequencies for which the slow rotation approximation is accurate enough) -- $f = 160$ Hz. In both panels one can see that if $\nu_n$ is calculated on the surface of the star, the highest frequency is in GR and all STT cases have lower frequencies. Contrary, if the nodal precession frequency is calculated at ISCO, the GR ones is the lowest and all STT cases have higher frequencies. In the right panel (for $\beta = -10$), one can see the non-monotonic behavior of the deviation from GR with the parameters in the theory, which we have already discussed for the previous figures.  The deviation in the left panel are below 2\%, and in the right below 10\%.

We have already mentioned that we are using the slow rotation approximation, which works quite accurately for rotational frequencies typical for the most of the observed pulsars. However, except for the last figure, we presented only the results for the static models in this paper. The reason for this is the small effect the rotation has on the results. In the results for the radius of the ISCO, for example, the rotation with frequency $f = 160$ Hz leads to shrinking of the radius of the orbit with only about 4\%. The deviations in the rest of the studied quantities are from the same magnitude. This makes us to believe that in the context of the EOS uncertainty and the three free parameters in the theory  the static case is a good enough approximation, even if one study slowly rotating models.

\section{Conclusion}

In this paper we studied the radius of ISCO, the orbital and the epicyclic frequency of particle moving on a circular orbit around neutron stars in massive scalar-tensor theory and in massive scalar-tensor theory with self-interaction. 
We found that the radius of ISCO in both theories is always bigger than the corresponding one in GR, and the orbital and the epicyclic frequencies are always lower compared to the GR ones. For the examined set of parameters we found the maximal deviation for ISCO in STT with self-interaction to be about 40\%. The maximal observed decrease in the orbital frequency is about 35\%, and in the maximal radial frequency about 50\%.

We conducted our numerical study in the so-called slow rotation approximation, which however is suitable for most of the observed rotation rates. We found that in this approximation for rotational frequency $f=160$ Hz (this is the highest rotation rate we have studied) the radius of ISCO decreases with about 4 \% compared to the static case in all studied cases (theories and combinations of parameters). The deviations for the rest of the studied quantities are from the same magnitude. Although this deviations are comparable with some of the deviations due to the modification of the theory, we tend to believe that even the static case is good approximation, which may allow us to constrain the parameters for which the higher deviations are observed. 

It is worth mentioning that for the case with lower value of beta we studied ($\beta = -10$) we observed a non-monotonic behavior in the deviation from GR with some of the parameters. This is direct result from the fact that the mass of the models at which ISCO appears changes with the parameters in the theory, and it may be lower or higher compared to the GR case, but it eventually converge to GR.

\section*{Acknowledgements}

KS, SY, and DD would like to thank for support by the COST Actions CA16214, CA16104 and CA15117.  KS is supported by the Bulgarian NSF Grant DM 18/4. DD would like to thank the European Social Fund, the Ministry of Science, Research and the Arts Baden-W\"{u}rttemberg for the support. DD is indebted to the Baden-W\"{u}rttemberg Stiftung for the financial support of this research project by the Eliteprogramme for Postdocs. DP and SY are supported partially by the Sofia University Grants No 80-10-73/2018 and No 3258/2017.


\bibliography{references}

\begin{thebibliography}{31}
\expandafter\ifx\csname natexlab\endcsname\relax\def\natexlab#1{#1}\fi
\expandafter\ifx\csname bibnamefont\endcsname\relax
  \def\bibnamefont#1{#1}\fi
\expandafter\ifx\csname bibfnamefont\endcsname\relax
  \def\bibfnamefont#1{#1}\fi
\expandafter\ifx\csname citenamefont\endcsname\relax
  \def\citenamefont#1{#1}\fi
\expandafter\ifx\csname url\endcsname\relax
  \def\url#1{\texttt{#1}}\fi
\expandafter\ifx\csname urlprefix\endcsname\relax\def\urlprefix{URL }\fi
\providecommand{\bibinfo}[2]{#2}
\providecommand{\eprint}[2][]{\url{#2}}

\bibitem[{\citenamefont{Ramazano{\u{g}}lu and
  Pretorius}(2016)}]{Ramazanoglu2016}
\bibinfo{author}{\bibfnamefont{F.~M.} \bibnamefont{Ramazano{\u{g}}lu}}
  \bibnamefont{and}
  \bibinfo{author}{\bibfnamefont{F.}~\bibnamefont{Pretorius}},
  \bibinfo{journal}{Physical Review D} \textbf{\bibinfo{volume}{93}},
  \bibinfo{pages}{064005} (\bibinfo{year}{2016}).

\bibitem[{\citenamefont{Popchev}(2015)}]{Popchev2015}
\bibinfo{author}{\bibfnamefont{D.}~\bibnamefont{Popchev}}, Master's thesis,
  \bibinfo{school}{University of Sofia} (\bibinfo{year}{2015}).

\bibitem[{\citenamefont{Yazadjiev et~al.}(2016)\citenamefont{Yazadjiev, Doneva,
  and Popchev}}]{Yazadjiev2016}
\bibinfo{author}{\bibfnamefont{S.~S.} \bibnamefont{Yazadjiev}},
  \bibinfo{author}{\bibfnamefont{D.~D.} \bibnamefont{Doneva}},
  \bibnamefont{and} \bibinfo{author}{\bibfnamefont{D.}~\bibnamefont{Popchev}},
  \bibinfo{journal}{Physical Review D} \textbf{\bibinfo{volume}{93}},
  \bibinfo{pages}{084038} (\bibinfo{year}{2016}).

\bibitem[{\citenamefont{Staykov et~al.}(2018)\citenamefont{Staykov, Popchev,
  Doneva, and Yazadjiev}}]{Staykov2018}
\bibinfo{author}{\bibfnamefont{K.~V.} \bibnamefont{Staykov}},
  \bibinfo{author}{\bibfnamefont{D.}~\bibnamefont{Popchev}},
  \bibinfo{author}{\bibfnamefont{D.~D.} \bibnamefont{Doneva}},
  \bibnamefont{and} \bibinfo{author}{\bibfnamefont{S.~S.}
  \bibnamefont{Yazadjiev}}, \bibinfo{journal}{The European Physical Journal C}
  \textbf{\bibinfo{volume}{78}}, \bibinfo{pages}{1} (\bibinfo{year}{2018}),
  ISSN \bibinfo{issn}{1434-6052},
  \urlprefix\url{http://dx.doi.org/10.1140/epjc/s10052-018-6064-x}.

\bibitem[{\citenamefont{Popchev et~al.}(2018)\citenamefont{Popchev, Staykov,
  Doneva, and Yazadjiev}}]{Popchev2018}
\bibinfo{author}{\bibfnamefont{D.}~\bibnamefont{Popchev}},
  \bibinfo{author}{\bibfnamefont{K.~V.} \bibnamefont{Staykov}},
  \bibinfo{author}{\bibfnamefont{D.~D.} \bibnamefont{Doneva}},
  \bibnamefont{and} \bibinfo{author}{\bibfnamefont{S.~S.}
  \bibnamefont{Yazadjiev}}, \bibinfo{journal}{arXiv e-prints}
  (\bibinfo{year}{2018}), \eprint{1812.00347},
  \urlprefix\url{http://adsabs.harvard.edu/abs/2018arXiv181200347P}.

\bibitem[{\citenamefont{Collaboration et~al.}(2018)\citenamefont{Collaboration,
  the Virgo~Collaboration, Abbott, Abbott, Abbott, Abraham, Acernese, Ackley,
  Adams, Adhikari et~al.}}]{Collaboration2018}
\bibinfo{author}{\bibfnamefont{T.~L.~S.} \bibnamefont{Collaboration}},
  \bibinfo{author}{\bibnamefont{the Virgo~Collaboration}},
  \bibinfo{author}{\bibfnamefont{B.~P.} \bibnamefont{Abbott}},
  \bibinfo{author}{\bibfnamefont{R.}~\bibnamefont{Abbott}},
  \bibinfo{author}{\bibfnamefont{T.~D.} \bibnamefont{Abbott}},
  \bibinfo{author}{\bibfnamefont{S.}~\bibnamefont{Abraham}},
  \bibinfo{author}{\bibfnamefont{F.}~\bibnamefont{Acernese}},
  \bibinfo{author}{\bibfnamefont{K.}~\bibnamefont{Ackley}},
  \bibinfo{author}{\bibfnamefont{C.}~\bibnamefont{Adams}},
  \bibinfo{author}{\bibfnamefont{R.~X.} \bibnamefont{Adhikari}},
  \bibnamefont{et~al.}, \bibinfo{journal}{arXiv e-prints}
  (\bibinfo{year}{2018}), \eprint{1811.12907}.

\bibitem[{\citenamefont{Abbott et~al.}(2017)\citenamefont{Abbott, Abbott,
  Abbott, Acernese, Ackley, Adams, Adams, Addesso, Adhikari, Adya
  et~al.}}]{Abbott2017}
\bibinfo{author}{\bibfnamefont{B.~P.} \bibnamefont{Abbott}},
  \bibinfo{author}{\bibfnamefont{R.}~\bibnamefont{Abbott}},
  \bibinfo{author}{\bibfnamefont{T.~D.} \bibnamefont{Abbott}},
  \bibinfo{author}{\bibfnamefont{F.}~\bibnamefont{Acernese}},
  \bibinfo{author}{\bibfnamefont{K.}~\bibnamefont{Ackley}},
  \bibinfo{author}{\bibfnamefont{C.}~\bibnamefont{Adams}},
  \bibinfo{author}{\bibfnamefont{T.}~\bibnamefont{Adams}},
  \bibinfo{author}{\bibfnamefont{P.}~\bibnamefont{Addesso}},
  \bibinfo{author}{\bibfnamefont{R.~X.} \bibnamefont{Adhikari}},
  \bibinfo{author}{\bibfnamefont{V.~B.} \bibnamefont{Adya}},
  \bibnamefont{et~al.}, \bibinfo{journal}{Physical Review Letters}
  \textbf{\bibinfo{volume}{119}}, \bibinfo{eid}{161101} (\bibinfo{year}{2017}),
  \eprint{1710.05832},
  \urlprefix\url{http://adsabs.harvard.edu/abs/2017PhRvL.119p1101A}.

\bibitem[{\citenamefont{van~der Klis~in}(2006)}]{Klis2006}
\bibinfo{author}{\bibfnamefont{M.}~\bibnamefont{van~der Klis~in}},
  \emph{\bibinfo{title}{Compact Stellar X-ray Sources (Cambridge
  Astrophysics)}} (\bibinfo{publisher}{Cambridge University Press},
  \bibinfo{year}{2006}), ISBN \bibinfo{isbn}{0521826594}.

\bibitem[{\citenamefont{Miller et~al.}(1998)\citenamefont{Miller, Lamb, and
  Psaltis}}]{Miller1998}
\bibinfo{author}{\bibfnamefont{M.~C.} \bibnamefont{Miller}},
  \bibinfo{author}{\bibfnamefont{F.~K.} \bibnamefont{Lamb}}, \bibnamefont{and}
  \bibinfo{author}{\bibfnamefont{D.}~\bibnamefont{Psaltis}},
  \bibinfo{journal}{\apj} \textbf{\bibinfo{volume}{508}}, \bibinfo{pages}{791}
  (\bibinfo{year}{1998}), \eprint{astro-ph/9609157},
  \urlprefix\url{http://adsabs.harvard.edu/abs/1998ApJ...508..791M}.

\bibitem[{\citenamefont{Stella and Vietri}(1999)}]{Stella1999}
\bibinfo{author}{\bibfnamefont{L.}~\bibnamefont{Stella}} \bibnamefont{and}
  \bibinfo{author}{\bibfnamefont{M.}~\bibnamefont{Vietri}},
  \bibinfo{journal}{Physical Review Letters} \textbf{\bibinfo{volume}{82}},
  \bibinfo{pages}{17} (\bibinfo{year}{1999}), \eprint{astro-ph/9812124},
  \urlprefix\url{http://adsabs.harvard.edu/abs/1999PhRvL..82...17S}.

\bibitem[{\citenamefont{Stella}(2001)}]{Stella2001}
\bibinfo{author}{\bibfnamefont{L.}~\bibnamefont{Stella}},
  \bibinfo{journal}{X-ray Astronomy: Stellar Endpoints, AGN, and the Diffuse
  X-ray Background} \textbf{\bibinfo{volume}{599}}, \bibinfo{pages}{365}
  (\bibinfo{year}{2001}), \eprint{astro-ph/0011395},
  \urlprefix\url{http://adsabs.harvard.edu/abs/2001AIPC..599..365S}.

\bibitem[{\citenamefont{Abramowicz et~al.}(2004)\citenamefont{Abramowicz,
  Kluzniak, Stuchlik, and Torok}}]{Abramowicz2004}
\bibinfo{author}{\bibfnamefont{M.~A.} \bibnamefont{Abramowicz}},
  \bibinfo{author}{\bibfnamefont{W.}~\bibnamefont{Kluzniak}},
  \bibinfo{author}{\bibfnamefont{Z.}~\bibnamefont{Stuchlik}}, \bibnamefont{and}
  \bibinfo{author}{\bibfnamefont{G.}~\bibnamefont{Torok}},
  \bibinfo{journal}{arXiv Astrophysics e-prints}  (\bibinfo{year}{2004}),
  \eprint{astro-ph/0401464},
  \urlprefix\url{http://adsabs.harvard.edu/abs/2004astro.ph..1464A}.

\bibitem[{\citenamefont{{Pappas}}(2012)}]{Pappas2012b}
\bibinfo{author}{\bibfnamefont{G.}~\bibnamefont{{Pappas}}},
  \bibinfo{journal}{\mnras} \textbf{\bibinfo{volume}{422}},
  \bibinfo{pages}{2581} (\bibinfo{year}{2012}).

\bibitem[{\citenamefont{Motta et~al.}(2014)\citenamefont{Motta, Belloni,
  Stella, Mu{\~n}oz-Darias, and Fender}}]{Motta2014}
\bibinfo{author}{\bibfnamefont{S.~E.} \bibnamefont{Motta}},
  \bibinfo{author}{\bibfnamefont{T.~M.} \bibnamefont{Belloni}},
  \bibinfo{author}{\bibfnamefont{L.}~\bibnamefont{Stella}},
  \bibinfo{author}{\bibfnamefont{T.}~\bibnamefont{Mu{\~n}oz-Darias}},
  \bibnamefont{and} \bibinfo{author}{\bibfnamefont{R.}~\bibnamefont{Fender}},
  \bibinfo{journal}{\mnras} \textbf{\bibinfo{volume}{437}},
  \bibinfo{pages}{2554} (\bibinfo{year}{2014}), \eprint{1309.3652},
  \urlprefix\url{http://adsabs.harvard.edu/abs/2014MNRAS.437.2554M}.

\bibitem[{\citenamefont{{Pappas} and {Sotiriou}}(2015)}]{Pappas2015c}
\bibinfo{author}{\bibfnamefont{G.}~\bibnamefont{{Pappas}}} \bibnamefont{and}
  \bibinfo{author}{\bibfnamefont{T.~P.} \bibnamefont{{Sotiriou}}},
  \bibinfo{journal}{\mnras} \textbf{\bibinfo{volume}{453}},
  \bibinfo{pages}{2862} (\bibinfo{year}{2015}), \eprint{1505.02882}.

\bibitem[{\citenamefont{{Maselli} et~al.}(2015)\citenamefont{{Maselli},
  {Gualtieri}, {Pani}, {Stella}, and {Ferrari}}}]{Maselli2015}
\bibinfo{author}{\bibfnamefont{A.}~\bibnamefont{{Maselli}}},
  \bibinfo{author}{\bibfnamefont{L.}~\bibnamefont{{Gualtieri}}},
  \bibinfo{author}{\bibfnamefont{P.}~\bibnamefont{{Pani}}},
  \bibinfo{author}{\bibfnamefont{L.}~\bibnamefont{{Stella}}}, \bibnamefont{and}
  \bibinfo{author}{\bibfnamefont{V.}~\bibnamefont{{Ferrari}}},
  \bibinfo{journal}{\apj} \textbf{\bibinfo{volume}{801}}, \bibinfo{eid}{115}
  (\bibinfo{year}{2015}), \eprint{1412.3473}.

\bibitem[{\citenamefont{{Staykov} et~al.}(2015)\citenamefont{{Staykov},
  {Doneva}, and {Yazadjiev}}}]{Staykov2015a}
\bibinfo{author}{\bibfnamefont{K.~V.} \bibnamefont{{Staykov}}},
  \bibinfo{author}{\bibfnamefont{D.~D.} \bibnamefont{{Doneva}}},
  \bibnamefont{and} \bibinfo{author}{\bibfnamefont{S.~S.}
  \bibnamefont{{Yazadjiev}}}, \bibinfo{journal}{European Physical Journal C}
  \textbf{\bibinfo{volume}{75}}, \bibinfo{eid}{607} (\bibinfo{year}{2015}),
  \eprint{1508.07790}.

\bibitem[{\citenamefont{{Rezzolla}
  et~al.}(2003{\natexlab{a}})\citenamefont{{Rezzolla}, {Yoshida}, {Maccarone},
  and {Zanotti}}}]{Rezzolla2003}
\bibinfo{author}{\bibfnamefont{L.}~\bibnamefont{{Rezzolla}}},
  \bibinfo{author}{\bibfnamefont{S.}~\bibnamefont{{Yoshida}}},
  \bibinfo{author}{\bibfnamefont{T.~J.} \bibnamefont{{Maccarone}}},
  \bibnamefont{and}
  \bibinfo{author}{\bibfnamefont{O.}~\bibnamefont{{Zanotti}}},
  \bibinfo{journal}{\mnras} \textbf{\bibinfo{volume}{344}},
  \bibinfo{pages}{L37} (\bibinfo{year}{2003}{\natexlab{a}}).

\bibitem[{\citenamefont{{Rezzolla}
  et~al.}(2003{\natexlab{b}})\citenamefont{{Rezzolla}, {Yoshida}, and
  {Zanotti}}}]{Rezzolla2003a}
\bibinfo{author}{\bibfnamefont{L.}~\bibnamefont{{Rezzolla}}},
  \bibinfo{author}{\bibfnamefont{S.}~\bibnamefont{{Yoshida}}},
  \bibnamefont{and}
  \bibinfo{author}{\bibfnamefont{O.}~\bibnamefont{{Zanotti}}},
  \bibinfo{journal}{\mnras} \textbf{\bibinfo{volume}{344}},
  \bibinfo{pages}{978} (\bibinfo{year}{2003}{\natexlab{b}}).

\bibitem[{\citenamefont{{Montero} et~al.}(2004)\citenamefont{{Montero},
  {Rezzolla}, and {Yoshida}}}]{Montero2004}
\bibinfo{author}{\bibfnamefont{P.~J.} \bibnamefont{{Montero}}},
  \bibinfo{author}{\bibfnamefont{L.}~\bibnamefont{{Rezzolla}}},
  \bibnamefont{and}
  \bibinfo{author}{\bibfnamefont{S.}~\bibnamefont{{Yoshida}}},
  \bibinfo{journal}{\mnras} \textbf{\bibinfo{volume}{354}},
  \bibinfo{pages}{1040} (\bibinfo{year}{2004}).

\bibitem[{\citenamefont{Fragile et~al.}(2016)\citenamefont{Fragile, Straub, and
  Blaes}}]{Fragile2016}
\bibinfo{author}{\bibfnamefont{P.~C.} \bibnamefont{Fragile}},
  \bibinfo{author}{\bibfnamefont{O.}~\bibnamefont{Straub}}, \bibnamefont{and}
  \bibinfo{author}{\bibfnamefont{O.}~\bibnamefont{Blaes}},
  \bibinfo{journal}{\mnras} \textbf{\bibinfo{volume}{461}},
  \bibinfo{pages}{1356} (\bibinfo{year}{2016}), \eprint{1602.08082},
  \urlprefix\url{http://adsabs.harvard.edu/abs/2016MNRAS.461.1356F}.

\bibitem[{\citenamefont{de~Avellar et~al.}(2018)\citenamefont{de~Avellar,
  Porth, Younsi, and Rezzolla}}]{Avellar2018}
\bibinfo{author}{\bibfnamefont{M.~G.~B.} \bibnamefont{de~Avellar}},
  \bibinfo{author}{\bibfnamefont{O.}~\bibnamefont{Porth}},
  \bibinfo{author}{\bibfnamefont{Z.}~\bibnamefont{Younsi}}, \bibnamefont{and}
  \bibinfo{author}{\bibfnamefont{L.}~\bibnamefont{Rezzolla}},
  \bibinfo{journal}{\mnras} \textbf{\bibinfo{volume}{474}},
  \bibinfo{pages}{3967} (\bibinfo{year}{2018}),
  \urlprefix\url{http://adsabs.harvard.edu/abs/2018MNRAS.474.3967D}.

\bibitem[{\citenamefont{Doneva and Yazadjiev}(2016)}]{Doneva:2016xmf}
\bibinfo{author}{\bibfnamefont{D.~D.} \bibnamefont{Doneva}} \bibnamefont{and}
  \bibinfo{author}{\bibfnamefont{S.~S.} \bibnamefont{Yazadjiev}},
  \bibinfo{journal}{JCAP} \textbf{\bibinfo{volume}{1611}}, \bibinfo{pages}{019}
  (\bibinfo{year}{2016}), \eprint{1607.03299}.

\bibitem[{\citenamefont{{Hartle}}(1967)}]{Hartle1967}
\bibinfo{author}{\bibfnamefont{J.~B.} \bibnamefont{{Hartle}}},
  \bibinfo{journal}{\apj} \textbf{\bibinfo{volume}{150}}, \bibinfo{pages}{1005}
  (\bibinfo{year}{1967}).

\bibitem[{\citenamefont{Yazadjiev et~al.}(2014)\citenamefont{Yazadjiev, Doneva,
  Kokkotas, and Staykov}}]{Yazadjiev2014}
\bibinfo{author}{\bibfnamefont{S.~S.} \bibnamefont{Yazadjiev}},
  \bibinfo{author}{\bibfnamefont{D.~D.} \bibnamefont{Doneva}},
  \bibinfo{author}{\bibfnamefont{K.~D.} \bibnamefont{Kokkotas}},
  \bibnamefont{and} \bibinfo{author}{\bibfnamefont{K.~V.}
  \bibnamefont{Staykov}}, \bibinfo{journal}{JCAP}
  \textbf{\bibinfo{volume}{1406}}, \bibinfo{pages}{003} (\bibinfo{year}{2014}).

\bibitem[{\citenamefont{{Ryan}}(1995)}]{Ryan1995}
\bibinfo{author}{\bibfnamefont{F.~D.} \bibnamefont{{Ryan}}},
  \bibinfo{journal}{\prd} \textbf{\bibinfo{volume}{52}}, \bibinfo{pages}{5707}
  (\bibinfo{year}{1995}).

\bibitem[{\citenamefont{{Shibata} and {Sasaki}}(1998)}]{Shibata1998}
\bibinfo{author}{\bibfnamefont{M.}~\bibnamefont{{Shibata}}} \bibnamefont{and}
  \bibinfo{author}{\bibfnamefont{M.}~\bibnamefont{{Sasaki}}},
  \bibinfo{journal}{\prd} \textbf{\bibinfo{volume}{58}}, \bibinfo{eid}{104011}
  (\bibinfo{year}{1998}), \eprint{gr-qc/9807046}.

\bibitem[{\citenamefont{{Pappas} and {Apostolatos}}(2012)}]{Pappas2012}
\bibinfo{author}{\bibfnamefont{G.}~\bibnamefont{{Pappas}}} \bibnamefont{and}
  \bibinfo{author}{\bibfnamefont{T.~A.} \bibnamefont{{Apostolatos}}},
  \bibinfo{journal}{Physical Review Letters} \textbf{\bibinfo{volume}{108}},
  \bibinfo{eid}{231104} (\bibinfo{year}{2012}).

\bibitem[{\citenamefont{{Akmal} et~al.}(1998)\citenamefont{{Akmal},
  {Pandharipande}, and {Ravenhall}}}]{AkmalPR}
\bibinfo{author}{\bibfnamefont{A.}~\bibnamefont{{Akmal}}},
  \bibinfo{author}{\bibfnamefont{V.~R.} \bibnamefont{{Pandharipande}}},
  \bibnamefont{and} \bibinfo{author}{\bibfnamefont{D.~G.}
  \bibnamefont{{Ravenhall}}}, \bibinfo{journal}{\prc}
  \textbf{\bibinfo{volume}{58}}, \bibinfo{pages}{1804} (\bibinfo{year}{1998}).

\bibitem[{\citenamefont{{Douchin} and {Haensel}}(2001)}]{Douchin2001}
\bibinfo{author}{\bibfnamefont{F.}~\bibnamefont{{Douchin}}} \bibnamefont{and}
  \bibinfo{author}{\bibfnamefont{P.}~\bibnamefont{{Haensel}}},
  \bibinfo{journal}{\aap} \textbf{\bibinfo{volume}{380}}, \bibinfo{pages}{151}
  (\bibinfo{year}{2001}).

\bibitem[{\citenamefont{{Read} et~al.}(2009)\citenamefont{{Read}, {Lackey},
  {Owen}, and {Friedman}}}]{Read2009}
\bibinfo{author}{\bibfnamefont{J.~S.} \bibnamefont{{Read}}},
  \bibinfo{author}{\bibfnamefont{B.~D.} \bibnamefont{{Lackey}}},
  \bibinfo{author}{\bibfnamefont{B.~J.} \bibnamefont{{Owen}}},
  \bibnamefont{and} \bibinfo{author}{\bibfnamefont{J.~L.}
  \bibnamefont{{Friedman}}}, \bibinfo{journal}{\prd}
  \textbf{\bibinfo{volume}{79}}, \bibinfo{eid}{124032} (\bibinfo{year}{2009}),
  \eprint{0812.2163}.

\end{thebibliography}

\end{document}